\documentclass[aps,pra,reprint,amssymb,amsmath]{revtex4-2}
\usepackage{graphicx}
\usepackage{amsthm}

\def\be{\begin{equation}}
\def\ee{\end{equation}}
\def\ber{\begin{eqnarray}}
\def\eer{\end{eqnarray}}
\def\bern{\begin{eqnarray*}}
\def\eern{\end{eqnarray*}}

\def\rv{\mathbf{r}}

\def\Gv{\mathbf{G}}
\def\kv{\mathbf{k}}

\def\0v{\mathbf{0}}
\def\1v{\mathbf{1}}
\def\2v{\mathbf{2}}
\def\3v{\mathbf{3}}

\def\pa{\partial}

\DeclareMathAlphabet\mathbfcal{OMS}{cmsy}{b}{n}

\newtheorem{theo}{Theorem}

\begin{document}

\title{Breakdown of the ionization potential theorem of density functional theory in mesoscopic systems}
\author{Vladimir~U.~Nazarov}
\affiliation{Moscow Institute of Physics and Technology (National Research University), Dolgoprudny, Russian Federation}
\email{nazarov.vu@mipt.ru}

\begin{abstract}
The IP-theorem of the Kohn-Sham (KS) density functional theory (DFT)  states that
the energy of the highest occupied molecular orbital (HOMO) $\epsilon_{HOMO}$ equals the negative of the first ionization potential (IP),
thus ascribing a physical meaning to one of the eigenvalues of the KS hamiltonian.
We scrutinize the fact that the validity of the IP-theorem relies critically on the electron density $n(\rv)$,
far from the system, to be determined by HOMO only,  behaving as $n(\rv)  \underset{r\to\infty}{\sim}  e^{- 2 \sqrt{-2 \epsilon_{HOMO}} r}$. 
While this behavior always holds for finite systems,  it does not hold for mesoscopic ones, such as quasi-two-dimensional (Q2D) electron gas or Q2D crystals. 
We show that this leads to the violation of the IP-theorem  for the latter class of systems.
This finding has a strong bearing on the role of the KS valence band with respect to the work-function problem in the  mesoscopic case.
Based  on  our results, 
we introduce a concept of the IP band structure as an observable alternative to its unphysical  KS counterpart.
A practical method of the determination of  IP band structure in terms of  DFT quantities is provided.
\end{abstract}

\maketitle

In the early years of quantum mechanics,  T. Koopmans had shown that, within the Hartree-Fock (HF) theory,
the energy of the highest occupied molecular orbital (HOMO), taken with minus sign, coincides with the  first ionization potential (IP) of the same system \cite{Koopman-34}. This result, known as Koopmans' theorem, plays an important role in quantum theory. Indeed, on the one hand, it ascribes a physical meaning to the HOMO eigenenergy, which, otherwise, is merely one of the Lagrange multipliers in the HF variational problem.  
On the other, from the practical point of view, 
it is extremely beneficial to be able to determine IP from a single HF calculation.

With the advance of the Kohn-Sham (KS) density functional theory (DFT) \cite{Kohn-65}, the question of the interpretation of orbital energies  has been, naturally, raised again, and it was answered by the IP-theorem \cite{Perdew-82,Perdew-83,Levy-84,Almbladh-85,Perdew-97}, which states
that HOMO energy and the negative of the first IP are equal  quantities (see, e.g., Ref.~\onlinecite{Kronik-20} for a recent review of this and related properties in the context of their spectroscopic significance).


A close examination of the IP-theorem reveals that a crucial condition of its validity is the electronic density $n(\rv)$, 
at large distance $\rv$ from the system, to be determined by HOMO only \cite{Perdew-97} (see also proof of our Theorem 1 below). Specifically, 
\begin{equation}
n(\rv)  \underset{r\to\infty}{\sim} e^{-2 \kappa r}, \
\kappa=  \sqrt{-2 \epsilon_{HOMO}} 
\label{asy}
\end{equation}
(we use atomic units unless indicated otherwise).
The property (\ref{asy}) holds for finite systems, 
since then HOMO is the orbital with the slowest falloff in vacuum.  Let us, however, consider a quasi-two-dimensional (Q2D) crystal. 
For the orbital $\phi_{\kv, n}(\rv)$ of the in-plane wave-vector $\kv$ within the first Brillouin zone and the band number $n$, we can write in the Laue representation \cite{Laue-31}
\begin{equation*}
\phi_{\kv, n}(\rv)= \sum\limits_{\Gv} u_{\kv,n,\Gv} (z) e^{i(\Gv+\kv)\cdot \rv_\|},
\end{equation*}
where $\Gv$ are the reciprocal in-plane lattice vectors. 
Therefore, at large $|z|$, where the KS potential is flat and zero 
(as taken relative to the vacuum level), we can write for the solution of the KS equation for a bound orbital
\begin{equation*}
\phi_{\kv, n}(\rv) \underset{ z \to \pm \infty} {\sim} \sum\limits_{\Gv} 
a^\pm_{\kv, n,\Gv} e^{-\kappa_{\kv,n,\Gv} |z| }e^{i(\Gv+\kv)\cdot \rv_\|},
\end{equation*}
where
\begin{equation}
\kappa_{\kv,n,\Gv} = \sqrt{(\kv+\Gv)^2-2 \epsilon_{\kv,n}},
\label{kappa}
\end{equation}
and $\epsilon_{\kv,n}$ is the orbital energy. Accordingly, 
\begin{equation}
n(\rv) \underset{|z|\to \infty}{\sim} e^{-2 \kappa_{min} |z|}, \
\kappa_{min} = \min\limits_{(\kv,n) \in occ}  \kappa_{\kv,n,\0v},
\label{kappamin}
\end{equation}
where the minimization is taken over all the occupied states. 
We see from Eqs.~(\ref{kappa}) and (\ref{kappamin}) that, for a Q2D crystal, and in contrast to the case of a finite system, the asymptotic behavior of the density  is not, in general, governed by $\epsilon_{HOMO}$,  the latter fact raising concerns regarding the validity of the IP-theorem.
This decisive difference between the finite and mesoscopic cases arises, of course, as a consequence of the fact that, for the latter, there exists no distance large compared with the system's size, since the size is infinite in one (or more) dimensions, while it is still possible to go far from the system along its microscopic dimension.

In this paper, with the focus on the two archetypal systems of mesoscopic physics, those of Q2D electron gas (Q2DEG) and graphene, we show that the IP-theorem breaks down, indeed.
We start by considering  Q2DEG  with one filled miniband and we demonstrate explicitly the IP-theorem's failure  within the exact exchange (EXX)  DFT. 
Next, we prove that the IP-theorem's violation within EXX leads necessarily to its violation within the exact DFT as well.
At the same time, within HF theory, we demonstrate that Koopmans'  theorem passes the test of Q2DEG.
Finally, from Q2DEG we turn to a system of a more practical significance, that of graphene, confirming  that all our findings for the former system  hold true for the latter one as well.

{\it Q2DEG with one filled miniband: KS DFT.}--
We consider electron gas, homogeneous in the $xy$-plane and confined in the $z$-direction by an external potential $v_{ext}(z)$. Due to the separation of variables in the KS equations, the in-plane and perpendicular motions  are independent. Below a (very moderate) threshold electron density value, only one orbital of the motion in the $z$-direction $\mu(z)$ gets occupied
\cite{Nazarov-16-2}, which produces Q2DEG with one filled miniband. For orbitals we can write 
\begin{equation}
\phi_\kv(\rv)= \Omega^{-1/2} e^{i \kv \cdot \rv_\|} \mu(z),
\label{DFTO}
\end{equation}
where $\Omega$ is the normalization area. 
$\mu(z)$ does not depend on $\kv$ and it satisfies the KS equation
\begin{equation}
\left[-\frac{1}{2}\frac{ d^2} {d z^2} 
+ v_{ext}(z)+v_H(z) +v_{xc}(z) \right] \mu(z) = 
\epsilon  \mu(z),
\label{KS}
\end{equation}
where $v_H(z)$ and $v_{xc}(z)$ are the Hartree and  exchange-correlation  potentials, respectively. 

\begin{figure} [h!] 
\includegraphics[width= \columnwidth, trim= 63 0 20 0, clip=true]{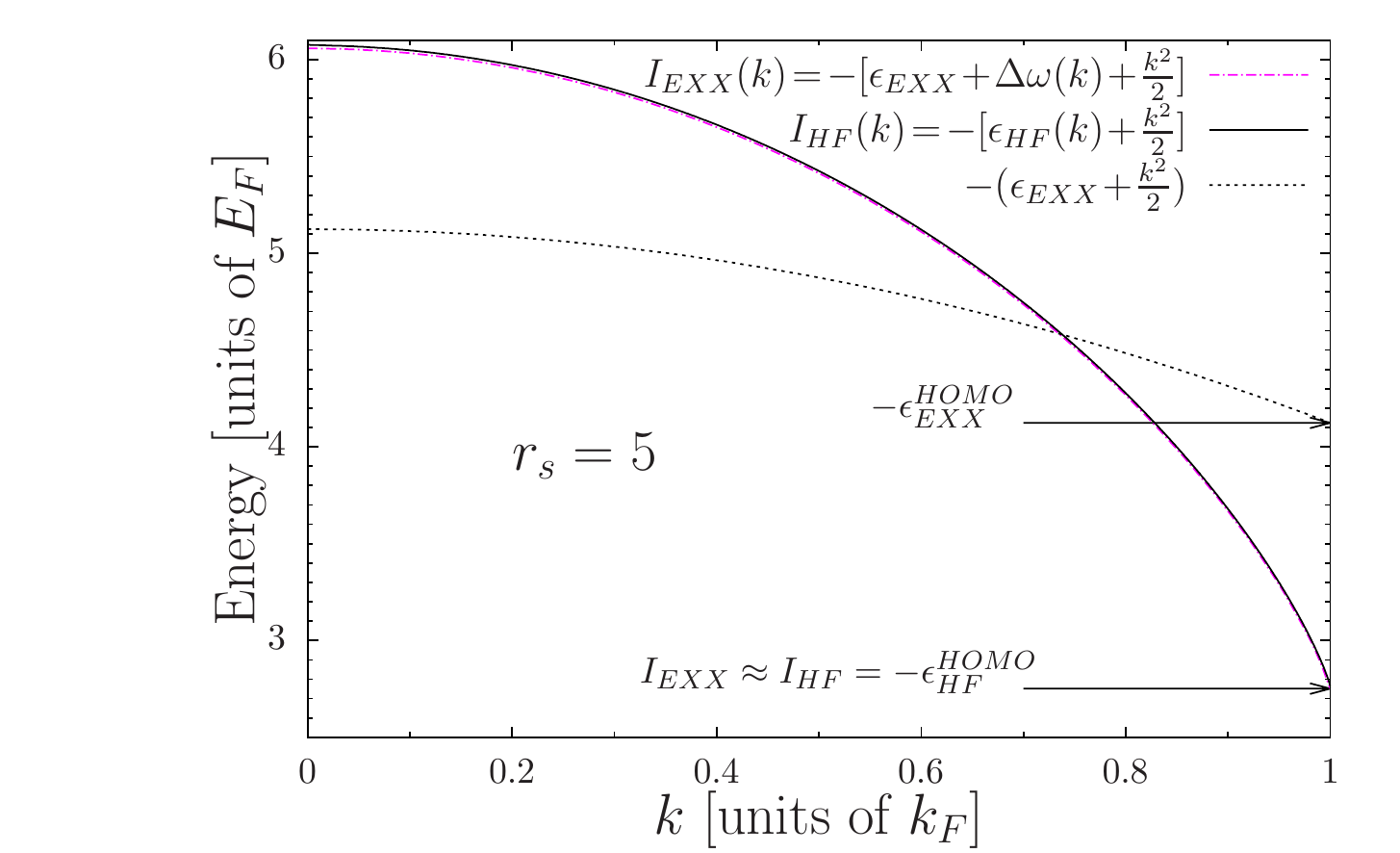}
\caption{\label{ipfig}
Ionization potential of Q2DEG with one filled miniband  versus the in-plane wave-vector $k$.
$I_{HF}(k)=-[\epsilon_{HF}(k)+\frac{k^2}{2}]$ in the HF theory, while $I_{EXX}(k)\ne -[\epsilon_{EXX}+\frac{k^2}{2}]$ in EXX DFT. Results for Q2DEG with the density parameter $r_s=5$ and the confining potential of the strictly 2D positively charged sheet in the $xy$-plane are shown.
The horizontal arrows indicate the minus HOMO energy of EXX and the 1st IP, the latter found as the minimal photon energy required to emit an electron. $I_{EXX}(k)$ and $I_{HF}(k)$ are almost indistinguishable from each other in the plot. 
}
\end{figure}

Q2DEG with one filled miniband is a unique extended system for which 
the exact-exchange (EXX) potential
(or optimized effective potential (OEP) \cite{Sharp-53,Talman-76}) 
is known analytically \cite{Nazarov-16-2,Nazarov-17},  
in both static (DFT) and dynamic [time-dependent DFT (TDDFT) \cite{Runge-84,Gross-85}]  cases. 
Restricting ourselves temporarily to EXX, 
we  can write 
\begin{equation}
\begin{split}
  v_x(z)   = -\int    \frac{F_2(k_F|z -  z'|)}{|z  - z'|}  |\mu (z')|^2 d z' ,
\end{split}
\label{main152}
\end{equation} 
where
$k_F$ is the in-plane Fermi radius and 
$F_2(u)$ is known explicitly in terms of special functions \cite{Nazarov-16-2}. Equations (\ref{KS}) and (\ref{main152}) are solved self-consistently, producing the EXX KS band structure of our system, which is determined by the $k$-independent eigenenergy $\epsilon$. 
The energy band of the 3D motion is $\epsilon_{\kv}=\epsilon+\frac{k^2}{2}$, 
the latter 
shown in Fig.~\ref{ipfig} with  dotted line. We note that, according to Eq.~(\ref{kappa}),  $\kappa_{\kv}=\sqrt{-2 \epsilon}$ is independent of $\kv$, 
with the result of {\it all} the occupied orbitals giving  the same contribution to the asymptotic behavior of the electron density, which, obviously, is inconsistent  with Eq.~(\ref{asy}).

\begin{figure} [h!] 
\includegraphics[width= \columnwidth, trim= 37 0 12 0, clip=true]{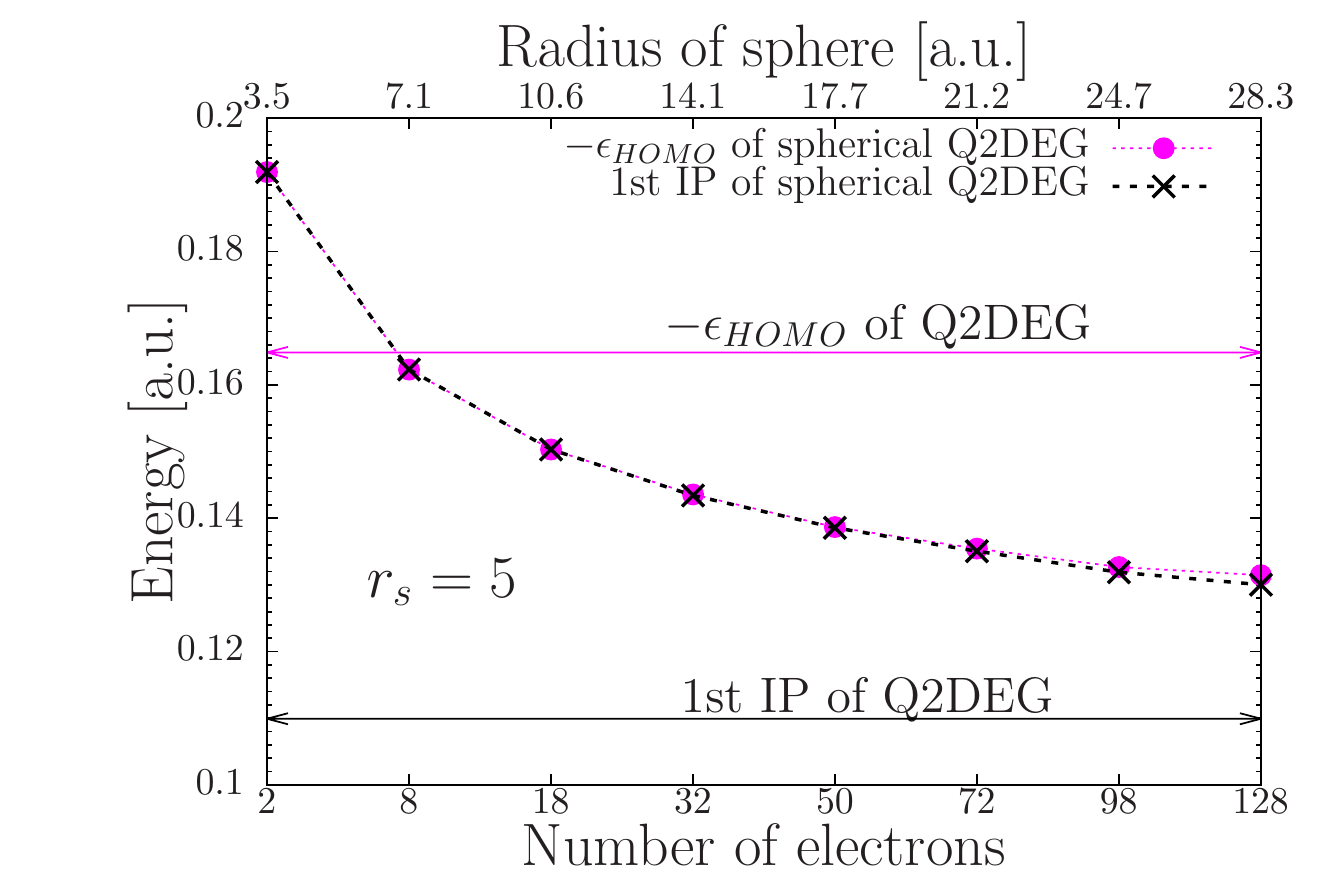}
\caption{\label{conv}
Minus HOMO energy
of the system of a finite number of electrons confined in the vicinity of the surface of a sphere (spherical Q2DEG) versus the number of electrons (filled circles). 
Corresponding IP's, obtained by Eqs.~(\ref{def3}) and (\ref{II}), do not differ from $-\epsilon_{HOMO}$ within the numerical accuracy (crosses).
Radius of the sphere is adjusted as to ensure a fixed surface particle density $n_s=1.27\times 10^{-2}$ a.u. ($r_s=5$).
Horizontal arrows show the minus HOMO energy and IP of the infinite system of Q2DEG, 
the two being {\em different quantities}.}
\end{figure}

In the case of a finite number of particles $N$, IP is usually  defined as
$I=E_{N-1}-E_N$,
where $E_N$ is the ground-state energy.
This definition, taken literally, becomes, however, meaningless if $N$ is infinite, since both $E_N$ and $E_{N-1}$ are infinite. Nor the definition of the IP as $-\frac{\pa E_N}{\pa N}$ with the continuous change of $N$ \cite{Perdew-82,Perdew-83,Perdew-97} can be used in the infinite case, since the function in the numerator is infinite.
To take use of the above definitions, one has to resort to the limiting procedure,
considering a sequence of finite systems mimicking the infinite one and expanding to the latter \cite{Perdew-17}.
There is, however, no guarantee that the sequence of IP$=-\epsilon_{HOMO}$ of finite systems converges to the 
$-\epsilon_{HOMO}$ of the infinite one (see Fig.~\ref{conv} and its discussion for the demonstration of the opposite).

To overcome the said difficulty, we identify IP with the minimal photon energy required to ionize a system,
which definition is equally applicable to  finite and infinite cases. 
Following approach of  Ref.~\cite{Nazarov-19}, we write down the probability, per unit time, for electron to be emitted by a weak external electric field into the final state $\phi_f$ as
\begin{equation}
P_f=\lim\limits_{t\to\infty} \frac{\langle \phi_f|\rho^{(2)}_0(t)+\rho^{(2)}_1(t)|\phi_f\rangle}{t},
\label{Pf}
\end{equation}
where $\rho(t)$ is the reduced one-particle density matrix (one-matrix), and the superscript and subscripts denote the order in the expansion of $\rho(t)$ in the series in powers of the applied  field and electron-electron  interaction, respectively.
In the former expansion, orders less than two do not contribute.
The latter expansion is done in the spirit of G\"{o}rling--Levy's adiabatic connection perturbation theory \cite{Gorling-94,Gorling-97}, and, in view of the further use at the level of  EXX, 
we restrict ourselves up to the first order in the interaction.

It was shown in Ref.~\cite{Nazarov-19} that the first term in Eq.~(\ref{Pf})  reproduces the conventional Fermi golden rule formula for the probability of electron emission 
\begin{equation}
\lim\limits_{t\to \infty} \! \! \frac{\langle \phi_f| \rho_0^{(2)}(t)|\phi_f\rangle}{t} \! = \! \frac{\pi }{2}
 \sum\limits_{i\in occ}  \! \!
|\langle \phi_f|v_s^{(1)}(\omega)|\phi_i\rangle|^2  \delta(\omega-\epsilon_f+\epsilon_i),
\label{ph0}
\end{equation}
where $v_s^{(1)}(\omega)$ is the time-dependent KS potential, $\omega$ is the frequency of the monocromatic applied field,
and the summation runs over all the occupied KS orbitals. 
Furthermore, while TDDFT, as well as the ground-state DFT, by the construction of the multiplicative potential $v_{xc}(\rv,t)$, delivers the physical particle  density in the form of the KS density $n(\rv,t)=\sum_{i\in occ} |\phi_i(\rv,t|^2$, 
the KS one-matrix
\begin{equation}
\rho_0(\rv,\rv',t)= \sum\limits_{i\in occ} \phi_i(\rv,t) \phi_i^*(\rv',t)
\end{equation}
does not provide  its physical counterpart $\rho(\rv,\rv',t)$ \cite[{\it cf}. Ref.][]{Casida-95}. 
This is the reason why, to consistently include the  interactions to the first order, we need to account for the second term in Eq.~(\ref{Pf}). This has been realized in Ref.~\cite{Nazarov-19} to the result
\begin{equation}
P_f \! = \! \! \!
 \sum\limits_{i\in occ}  \! \!
\left[\frac{\pi }{2} |\langle \phi_f|v_s^{(1)}(\omega)|\phi_i\rangle|^2  \! + \! \Delta A_{f i} \right] \delta(\omega-\epsilon_f+\epsilon_i +\Delta \omega_i),
\label{ph1}
\end{equation}
where $\Delta A_{f i}$ and $\Delta \omega_i$ are the interaction-caused shifts in the transitions strengths and the IPs from the corresponding orbitals, respectively, the latter  given by
\begin{equation}
\Delta \omega_i \! = \!  -\langle\phi_i|v_x^{(0)} |\phi_i\rangle
  -  \!   \int  \!  \! \rho_0^{(0)}(\rv,\rv')   \frac{\phi_i^*(\rv) \phi_i(\rv')}{|\rv-\rv'|}d\rv d\rv' .
 \label{def3}
\end{equation}
Ensuring the consistent inclusion of  interactions to the first order, Eq.~(\ref{ph1}) replaces Eq.~(\ref{ph0}) in the TDDFT-based theory of photoemission at the level of EXX, yielding the IP from the $\phi_i$ orbital as 
\begin{equation}
I_i=-(\epsilon_i+\Delta \omega_i).
\label{II}
\end{equation}
 
While in our example $\epsilon$ is $k$-indipendent,  $\Delta \omega(k)$ does depend on $k$. 
In Fig.~\ref{ipfig},  for Q2DEG with one filled miniband, the ionization potential $I_{EXX}(k)$ from the orbital with the in-plane wave-vector $k$, obtained with  the use of Eqs.~(\ref{def3}) and (\ref{II}), is plotted in the dashed-dotted line.
The lowest IP, which is the minimal photon energy needed to ionize the system, and $-\epsilon_{HOMO}=-(\epsilon+\frac{k_F^2}{2})$ are shown by horizontal arrows, and they are, by far, different quantities (see Appendix~\ref{AP3} for further particulars). 

{\it Q2DEG with one filled miniband:  HF theory.}--
It is impossible to satisfy HF equations with the orbitals of Eq.~(\ref{DFTO}) with $\mu(z)$ independent of $k$ ({\it cf.} Ref.~\cite{Luo-12}). Instead, we write the orbitals as
\begin{equation}
\phi_\kv(\rv)= \Omega^{-1/2} e^{i \kv \cdot \rv_\|} \mu_k(z), 
\label{HFO}
\end{equation}
which, after the substitution into HF equations, leads to
\begin{widetext}
\begin{equation}
\left[-\frac{1}{2}\frac{ d^2} {d z^2} 
+ v_{ext}(z)+v_H(z) \right] \mu_k(z) 
- \int \Theta(k_F-k')  k'   \mu^*_{k'} (z')  \mu_k(z') \mu_{k'}(z) 
H(k,k',|z-z'|)   d k'  d z'= 
 \epsilon_k  \mu_k(z),
\label{HFE}
\end{equation}
\end{widetext}
where $\Theta(k)$ is the Heaviside's step function and
\begin{equation}
H(k,k',u)= \frac{1}{2\pi} \int\limits_0^{2\pi}
\frac{e^{-u \sqrt{k^2+k'^2-2 k k' \cos \phi} }}{\sqrt{k^2+k'^2-2 k k' \cos \phi}} d \phi.
\label{Hfun}
\end{equation}
It can be seen from Eq.~(\ref{HFE}) that $\mu_k(z)$ do really depend on $k$, with the consequence that the perpendicular and in-plane motions in Q2DEG couple in the HF theory, the system's uniformity in the $xy$-plane notwithstanding.
In contrast to DFT, minibands are not flat any more.

We solve Eqs.~(\ref{HFE}) self-consistently, producing the HF band structure, which is plotted in Fig.~\ref{ipfig} with solid line. Remarkably, $I_{HF}(k)=-[\epsilon_{HF}(k)+\frac{k^2}{2}]$ is almost indistinguishable from 
$I_{EXX}(k)=-[\epsilon_{EXX}+\Delta \omega(k)+\frac{k^2}{2}]$, while them both  are very different from $-[\epsilon_{EXX}+\frac{k^2}{2}]$. 
The explanation of this is highly instructive: 
Unlike $v_{xc}(\rv,t)$ in (TD)DFT, the Fock {\em nonlocal} operator in the (TD)HF theory nullifies not only $n_1(\rv,t)$, but also $\rho_1(\rv,\rv',t)$. Therefore, in HF theory, $\rho=\rho_0$ up to the {\em first}, rather than to the zeroth, order in the interaction \cite[][{\it cf}. Ref.~\cite{Moller-34}]{Nazarov-19}, 
which results in the vanishing of the second term in Eq.~(\ref{Pf}). 
The latter, in its turn, leads to the validity of the Fermi golden rule (\ref{ph0}), rather than Eq.~(\ref{ph1}), within HF theory,
thus making  the IP equal to the minus eigenenergy of the corresponding level
\footnote{The same can be also seen by substituting the nonlocal Fock operator in place of the multiplicative potential $v_x^{(0)}$ in Eq.~(\ref{def3}), 
which results in  $\Delta \omega_i=0$.}.

It will be important for us that, for finite systems, the IP-theorem holds not only in exact DFT, but in EXX as well
\begin{theo}
\label{th1}
For a finite system  within EXX DFT, the energy shift $\Delta \omega_i$ of Eq.~(\ref{def3}) is zero for HOMO,
thus ensuring the equality of the first IP to the minus HOMO energy.
\end{theo}
Theorem \ref{th1} is proved  in Appendix \ref{AP1}. The same fact was earlier demonstrated numerically  in calculations for atoms \cite{Nazarov-19}.

We are now faced with a fundamental question: Is the violation of the IP-theorem, which  we have demonstrated for Q2DEG, pertinent to the EXX theory, or the same is also the case in the exact DFT? In other words, would the inclusion of correlations lead to the restoration of the equality between the IP and the minus HOMO energy? 
While this possibility looks unlikely from the outset, considering that, according to Theorem \ref{th1}, 
for finite systems the use of EXX does not break the IP-theorem, 
in Appendix~\ref{AP2} we prove that our results lead, necessarily, to the IP-theorem's violation within the exact DFT as well. 

We proceed by addressing a view of an infinite system  as a limit of the sequence of expanding finite ones, 
which has been used as a justification of the IP-theorem's validity in the infinite case
\cite{Perdew-17}.  To this end, we consider a sequence of spheres with electrons confined near the surface (spherical Q2DEG), simultaneously increasing the radius of a sphere and the number of electrons, while keeping the surface particle density fixed. In Fig.~\ref{conv} we follow the evolution of the $-\epsilon_{HOMO}$, which, for finite systems and within the numerical accuracy, coincides with IP, and we observe its tendency towards the IP of the infinite Q2DEG (physical quantity), rather than to the  $-\epsilon_{HOMO}$ value of the latter (unphysical quantity). We, therefore, conclude that the limiting procedure does not
preclude the violation of the IP-theorem in the infinite case. On the other hand, these results indicate that the above limiting procedure is, in principle,  legitimate for the determination of IP of an infinite system. This procedure is, however, absolutely impractical: For real materials, we cannot afford solving the KS problem for a sequence of clusters of increasing sizes.
Therefore, when having found  KS band-structure within the extended system setup (as it is being done routinely), we should be warned that its IP is not, generally speaking, given by  $-\epsilon_{HOMO}$.
We note, that there should be no surprise in the validity of the IP-theorem for a system of a finite $N$, {\it regardless of its magnitude},
and the theorem's invalidity for infinite $N$: In the former case, the $r\to \infty$ limit is taken, keeping $N$ finite, then Eq.~(\ref{asy}) holds.
In the latter, the limit $N\to \infty$ is taken first, then Eq.~(\ref{kappamin}) holds as $r\to\infty$. The two limits do not commute.

For thin jellium slabs, Luo {\it et al.}  \cite{Luo-12} have found
that  HF and EXX DFT energy band structures close to the Fermi surface are 
entirely different.  While this is in full agreement with our results, we have shown that the agreement between HF and EXX DFT  is restored if physical IP-band-structure rather than the unphysical KS one is used in the comparison of the two theories. 
Furthermore, for slabs of increasing thickness $d$, Luo {\it et al.}  were finding improving agreement between  EXX and HF theories \cite{Luo-12}.
This has the following qualitative explanation: For larger $d$, the number of minibands grows, while $E_F=\epsilon_n+\frac{k_{F,n}^2}{2}$ for all $n$,
where $n$ is the miniband's index and $k_{F,n}$ is its Fermi radius.
The extent of the density outside the slab is determined by $\kappa_{min}=\sqrt{-2 \epsilon_h}$, where $n=h$ is the index of the highest lying miniband.
Since $\epsilon_{HOMO}=\epsilon_h+\frac{k_{F,h}^2}{2}$, $k_{F,h}$ is the minimal among $k_{F,n}$, and we see that the difference between $\epsilon_{HOMO}$ and $\epsilon_h$ decreases with the increasing $d$, resulting in the density extent being more and more determined by $\epsilon_{HOMO}$. 
From this we conclude that our results cannot be transferred to semi-infinite crystals
({\it cf}. Ref.~\onlinecite{Almbladh-85}, where the validity of the IP-theorem for solid surfaces was asserted).

\begin{figure} [h!] 
\includegraphics[width=  \columnwidth, trim= 32 0 0 0, clip=true]{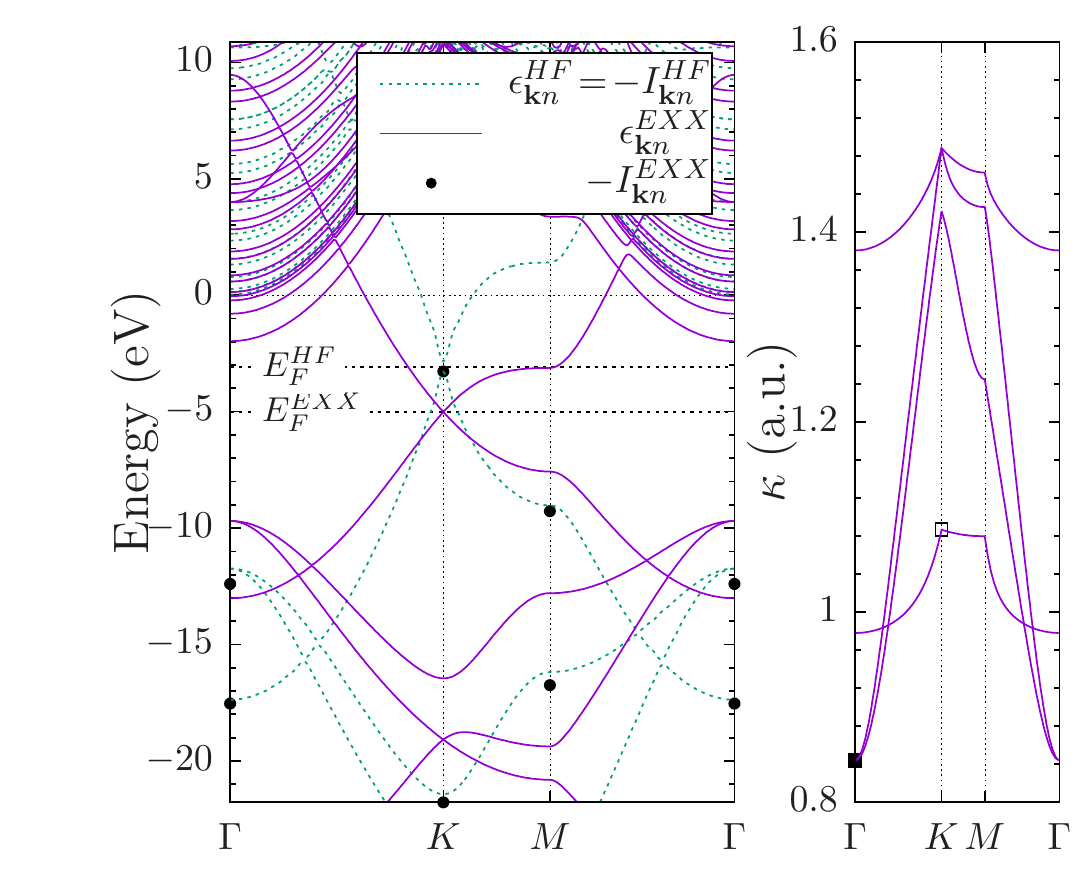}  
\caption{\label{graphene}
Left: Band structure of the monolayer graphene obtained within EXX DFT (solid line) and HF theory (dashed line), with zero energy at the vacuum level.
Solid circles show IP at symmetry points calculated  by Eqs.~(\ref{def3}) and (\ref{II}).
Right: $\kappa_{\kv,n,\0v}$ of Eq.~(\ref{kappa}).
Solid square marks $\kappa_{min}$ of Eq.~(\ref{kappamin}), demonstrating that the slowest decaying occupied orbital is not that of HOMO, the energies of the former and the latter (open square) lying at  $\Gamma$ and $K$ points, respectively. 
}
\end{figure}

{\it Graphene}.-- In Fig.~\ref{graphene}, we further illustrate our results  for the system of the pristine monolayer graphene.
In the left panel, we show graphene's band structure, calculated within  EXX DFT and HF theory. 
In the conceptual agreement with results for Q2DEG, HF and  EXX  band structures differ from each other significantly, as a manifestation of the fact  that the former is, while the latter is not, the IP band structure in the corresponding approximations. The agreement between the two theories is restored after the EXX IP is obtained with Eqs.~(\ref{def3}) and (\ref{II}) (shown  with solid circles for symmetry points). In order to illustrate that HOMO of graphene is not the slowest decaying, with the distance from the layer, occupied orbital,  in the right panel of Fig.~\ref{graphene} we plot $\kappa(\kv,n,\0v)$ of Eq.~(\ref{kappa}) versus $\kv$ for the four highest valence bands $n$
\footnote{Our calculations for graphene have been conducted with the all-electron full-potential linearized augmented-plane wave code Elk http://elk.sourceforge.net.}.


We note that the IP-theorem has been challenged in the literature before \cite{Kleinman-97,Kleinman-97-2}. The discussion, however, was conducted regarding finite systems, for which the validity of the theorem has been eventually reconfirmed  \cite{Perdew-97}.

{\it In conclusions}, we have addressed the problem of the validity of the ionization potential theorem of the density functional theory
in the case of a mesoscopic system - a system microscopic in one dimension and macroscopic in others.
We have shown that the IP-theorem, being true for systems of finite number of particles, breaks down in the mesoscopic case.
At the same time, we found that the Koopmans' theorem of the Hartree-Fock theory holds independently on the type of a system. 

We have traced the origin of this fundamental difference to the fact that the physical reduced density matrix, which includes all the information on the ionization process, coincides, to within the first order in the interaction,  with the HF density matrix, but not with the corresponding KS density matrix. 
Our findings suggest that  the work function of a mesoscopic system cannot, in general, be determined from the KS band structure, which is inherent to DFT itself rather than to the inaccuracies of specific approximations.

At the same time, we have shown an avenue to the consistent construction of work-functions of mesoscopic systems
from results obtained with DFT calculations. This is the use of the IP-band-structure, which is an observable physical quantity, rather than the unphysical KS one.
We have proposed a method of relating the two kinds of  band structures.
A remarkable agreement between the ionization potentials within the Hartree-Fock and the exact exchange density functional theory has been found, strongly supporting the promise of the proposed method. 
A way to the further advancement of the theory, which is the inclusion of correlations in the first place, 
can be clearly seen in the  construction of the physical reduced density-matrix 
as opposed to the use of its Kohn-Sham counterpart.

\acknowledgments
This work was supported by
Russian Foundation for Basic Research and the Ministry of Science and Technology of Taiwan  (Grant no. 21-51-52001).

%

\appendix

\onecolumngrid
\section{Proof of Theorem \ref{th1}.}
\label{AP1}

The EXX potential $v_x(\rv)$ satisfies the optimized effective potential (OEP) equation
\cite{Sharp-53,Talman-76}
\begin{equation}
\int \chi_s(\rv,\rv') v_x(\rv') d\rv' = - 2
\sum\limits_{\substack{\alpha, \in occ\\ \beta\in unocc}} 
\int \frac{\phi_\alpha^*(\rv'') \phi_\alpha(\rv)  \rho_0(\rv'',\rv')  \phi_\beta(\rv') \phi_\beta^*(\rv)}{|\rv'-\rv''| (\epsilon_\alpha-\epsilon_\beta)} d\rv' d\rv'',
\label{vx3}
\end{equation}
where $\chi_s(\rv,\rv') $ is the KS density response function given by Lindhard formula
\begin{equation}
\chi_s(\rv,\rv')    =  2 
\sum\limits_{\substack{\alpha\in occ \\ \beta\in unocc}}   
\frac{\phi_\alpha(\rv) \phi_\beta^*(\rv)  \phi_\beta(\rv') \phi_\alpha^*(\rv')}{\epsilon_\alpha - \epsilon_\beta} ,
\label{chi_sw0}
\end{equation}
$\alpha,\beta$ numerating the KS orbitals. 
Introducing the notation
\begin{equation}
f_\alpha(\rv,\rv')= \sum\limits_{\beta\in unocc} \frac{\phi_\beta^*(\rv) \phi_\beta(\rv')}{\epsilon_\alpha - \epsilon_\beta} 
=\sum\limits_{\beta\in unocc} \frac{\phi_\beta^*(\rv) \phi_\beta(\rv')}{\epsilon_\alpha +i\eta - \epsilon_\beta} =(\epsilon_\alpha+i \eta-\hat{h}_s)^{-1} \delta(\rv-\rv')-
\sum\limits_{\beta\in occ} \frac{\phi_\beta^*(\rv) \phi_\beta(\rv')}{\epsilon_\alpha +i \eta- \epsilon_\beta},
\label{feq}
\end{equation}
where $\hat{h}_s$ is the KS Hamiltonian and $\eta$ is an infinitesimal positive, we can write by Eqs.~(\ref{vx3})
and (\ref{chi_sw0})
\begin{equation}
\sum\limits_{\substack{\alpha\in occ }}   
\phi_\alpha(\rv)  \int f_\alpha(\rv,\rv') \phi_\alpha^*(\rv') v_x(\rv') d\rv' = 
- 
\sum\limits_{\alpha \in occ} \phi_\alpha(\rv) 
\int f_\alpha(\rv,\rv') \frac{\phi_\alpha^*(\rv'') \rho_0(\rv'',\rv')  } {|\rv'-\rv''|} d\rv' d\rv''.
\label{oep2}
\end{equation}
In Eq.~(\ref{feq}) the orthonormality and  completeness of the set of the orbitals has been used.
Together, Eqs.~(\ref{oep2}) and (\ref{feq}) give
\begin{equation}
\begin{split}
\sum\limits_{\substack{\alpha\in occ }}   
\phi_\alpha(\rv)  (\epsilon_\alpha+i \eta-\hat{h}_s)^{-1} \phi_\alpha^*(\rv) v_x(\rv)  
-\sum\limits_{\substack{\alpha,\beta\in occ }}   
\phi_\alpha(\rv)  \phi_\beta^*(\rv)  \int 
\frac{\phi_\beta(\rv')\phi_\alpha^*(\rv') v_x(\rv') }{\epsilon_\alpha +i \eta- \epsilon_\beta} d\rv' = \\
-
\sum\limits_{\alpha, \beta\in occ} \phi_\alpha(\rv) 
\int \phi_\alpha^*(\rv') \phi_\beta(\rv')(\epsilon_\alpha+i \eta-\hat{h}_s)^{-1}  \frac{\phi_\beta^*(\rv)} {|\rv-\rv'|}  d\rv' 
+
\sum\limits_{\alpha,\beta\in occ} \phi_\alpha(\rv) \phi_\beta^*(\rv) 
\int \frac{\phi_\alpha^*(\rv'') \rho_0(\rv'',\rv') \phi_\beta(\rv') } {|\rv'-\rv''| (\epsilon_\alpha +i \eta- \epsilon_\beta)} 
 d\rv' d\rv''.
\end{split}
\label{oep3}
\end{equation}

In the case of a finite system,  the HOMO orbital, which we denote by $\phi_{h}(\rv)$, dominates all the others at asymptotically large $r$
\cite{Perdew-97}.
We first consider the case of a non-degenerate $\epsilon_h$. Then, 
\begin{equation}
\begin{split} 
&\phi_h(\rv)  (\epsilon_h+i \eta-\hat{h}_s)^{-1} \phi_h^*(\rv) v_x(\rv)  
-  
|\phi_h(\rv)|^2  \int 
\frac{|\phi_h(\rv')|^2 v_x(\rv') }{i \eta} d\rv' 
\underset{r\to \infty} = \\
&-
\sum\limits_{ \beta\in occ} \phi_h(\rv) 
\int \phi_h^*(\rv') \phi_\beta(\rv')(\epsilon_h+i \eta-\hat{h}_s)^{-1}  \frac{\phi_\beta^*(\rv)} {|\rv-\rv'|}  d\rv' 
+
|\phi_h(\rv)|^2 
\int \frac{\phi_h^*(\rv'') \rho_0(\rv'',\rv') \phi_h(\rv') } {i \eta |\rv'-\rv''| } 
 d\rv' d\rv'',
\end{split}
\end{equation}
and canceling by $\phi_h(\rv)$,
\begin{equation}
\begin{split} 
&  (\epsilon_h+i \eta-\hat{h}_s)^{-1} \phi_h^*(\rv) v_x(\rv)  
-  
\phi_h^*(\rv)  \int 
\frac{|\phi_h(\rv')|^2 v_x(\rv') }{i \eta} d\rv' 
\underset{r\to \infty} = \\
&-
\sum\limits_{ \beta\in occ}  
\int \phi_h^*(\rv') \phi_\beta(\rv')(\epsilon_h+i \eta-\hat{h}_s)^{-1}  \frac{\phi_\beta^*(\rv)} {|\rv-\rv'|}  d\rv' 
+
 \phi_h^*(\rv) 
\int \frac{\phi_h^*(\rv'') \rho_0(\rv'',\rv') \phi_h(\rv') } {i \eta |\rv'-\rv''| } 
 d\rv' d\rv'',
\end{split}
\label{oep11}
\end{equation}

Since $\epsilon_h+i \eta-\hat{h}_s$ is a local operator, we can apply it on both sides of Eq.~(\ref{oep11}) at asymptotically large $r$. This gives
\begin{equation}
\begin{split} 
&   \phi_h^*(\rv) v_x(\rv)  
-  
\phi_h^*(\rv)  \int 
|\phi_h(\rv')|^2 v_x(\rv') d\rv' 
\underset{r\to \infty} = \\
&-
\sum\limits_{ \beta\in occ}  
\int \phi_h(\rv') \phi_\beta(\rv')  \frac{\phi_\beta^*(\rv)} {|\rv-\rv'|}  d\rv'
+
 \phi_h^*(\rv) 
\int \frac{\phi_h^*(\rv'') \rho_0(\rv'',\rv') \phi_h(\rv') } {|\rv'-\rv''| } 
 d\rv' d\rv'',
\end{split}
\label{oep12}
\end{equation}
which can  be written as
\begin{equation}
\begin{split} 
   \phi_h^*(\rv) v_x(\rv)  
-  
\phi_h^*(\rv)  \int 
|\phi_h(\rv')|^2 v_x(\rv') d\rv' 
\underset{r\to \infty} = 
-
  \frac{\phi_h^*(\rv)} {r} 
+
 \phi_h^*(\rv) 
\int \frac{\phi_h^*(\rv'') \rho_0(\rv',\rv'') \phi_h(\rv') } {|\rv'-\rv''| } 
 d\rv' d\rv'',
\end{split}
\label{oep13}
\end{equation}
where, in the 1st term on the RHS of  Eq.~(\ref{oep12}) we have accounted for $r$ being large. Canceling by $\phi_h^*(\rv)$, we have
\begin{equation}
v_x(\rv)  
-  
 \int 
|\phi_h(\rv')|^2 v_x(\rv') d\rv' 
\underset{r\to \infty} = 
-
  \frac{1} {r} 
+
\int \frac{\phi_h^*(\rv'') \rho_0(\rv'',\rv') \phi_h(\rv') } {|\rv'-\rv''| } 
 d\rv' d\rv''.
\label{oep14}
\end{equation}
It follows from Eq.~(\ref{oep14}) that
\begin{align}
&v_x(\rv)  = - \frac{1} {r} +C  \ \ \text{(a well known result),} \\
&-  
 \int 
|\phi_h(\rv')|^2 v_x(\rv') d\rv' 
= 
\int \frac{\phi_h^*(\rv'') \rho_0(\rv'',\rv') \phi_h(\rv') } {|\rv'-\rv''| } 
 d\rv' d\rv'' -C,
 \label{resff}
\end{align}
where $C$ is a constant. Setting $C=0$ \cite{Levy-84}, we conclude the proof of  Theorem \ref{th1} in the case of a non-degenerate HOMO.

If  HOMO is degenerate, we arrive at the same result by the same derivation by  taking account of the symmetries which cause the degeneracy. Then both sides in Eq.~(\ref{resff}) multiply by the order of the degeneracy, which leads to the same result.

\section{IP-theorem's violation: Extension from EXX to exact DFT.}
\label{AP2}

\begin{figure} [h!] 
\includegraphics[width= 0.5 \columnwidth, trim= 0 0 20 0, clip=true]{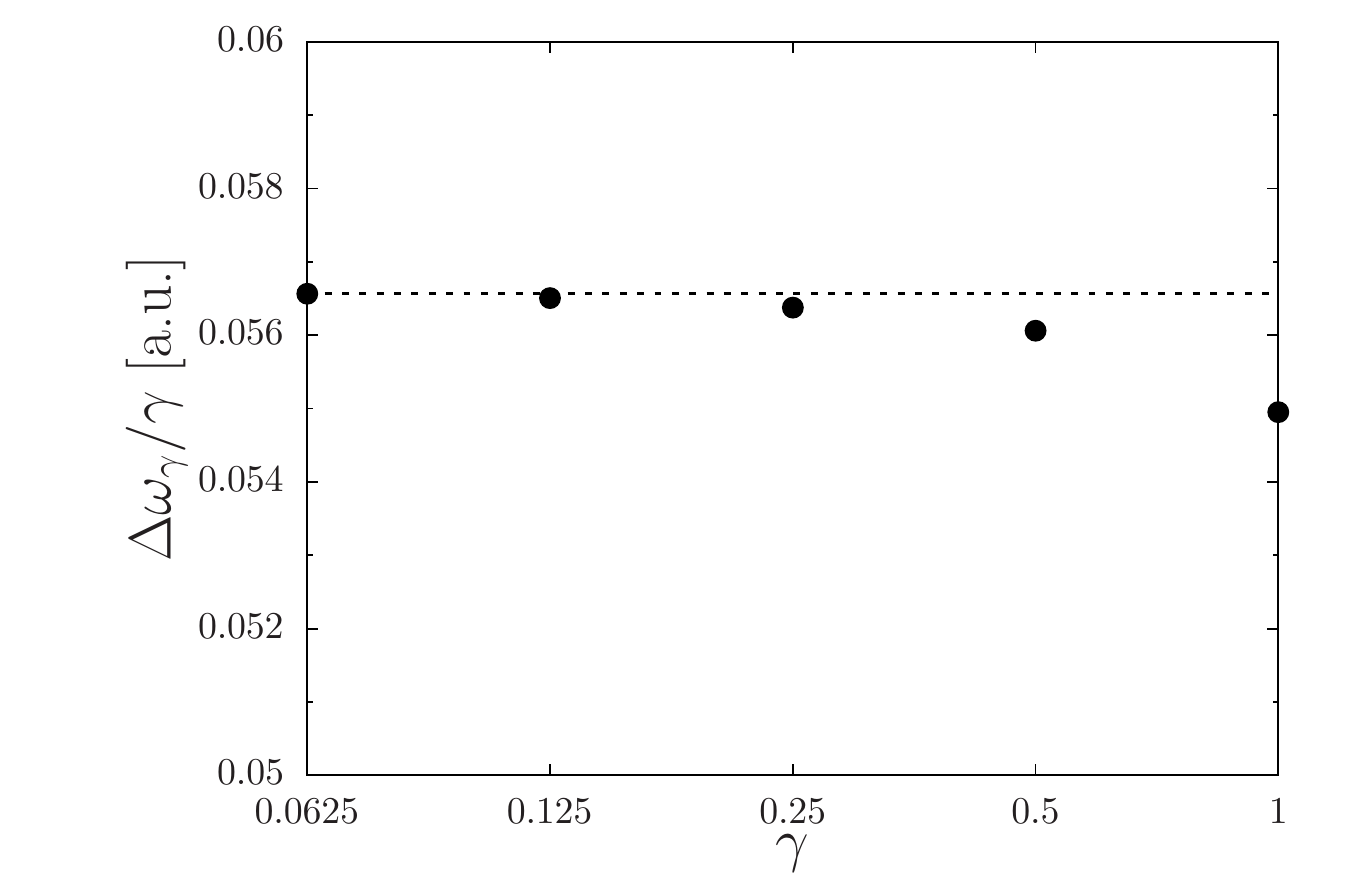}
\caption{\label{gamma}
Convergence of the ratio of $\Delta \omega_{HOMO,\gamma}$ to $\gamma$ to a constant at $\gamma\to 0$ 
is in contradiction to Eq.~(\ref{IE1}).
}
\end{figure}

Let us consider the scaled many-body Hamiltonian
\begin{equation}
\hat{H}_\gamma= \sum\limits_{i=1}^N \left[-\frac{1}{2} \Delta_i +v_{ext}(\rv_i)\right]
+\frac{1}{2} \sum\limits_{i\ne j}^N \frac{\gamma}{|\rv_i-\rv_j|}.
\label{sh}
\end{equation}
We are thinking of the Hamiltonian (\ref{sh}) as 'physical' with the modified Coulomb interaction,
and we want to construct the DFT corresponding to this modified  many-body problem. 
The KS potential, which depends on $\gamma$, can be written as
\begin{equation}
v_s(\rv;\gamma)=v_{ext}(\rv) + v_H(\rv;\gamma)+v_{xc}(\rv;\gamma).
\end{equation}
Obviously, the validity of the IP-theorem should not depend on the particular value of $\gamma$:
 If the theorem is valid, it should be valid for any $\gamma\in (0,1]$. Let us consider the limit $\gamma\to 0$.
 It is easy to realize that
\begin{equation}
v_s(\rv;\gamma)=v_{ext}(\rv) + v_H(\rv;\gamma)+v_x(\rv;\gamma) +O[\gamma]^2,
\label{AAA}
\end{equation}
where $v_H$ and $v_x$ are the Hartree and EXX potentials, respectively. 
The validity of Eq.~(\ref{AAA}) can be verified, e.g, with the G\"{o}rling--Levy's adiabatic connection perturbation procedure \cite{Gorling-94}.
Similarly,
\begin{align}
&\epsilon_{HOMO,\gamma}=\epsilon^{EXX}_{HOMO,\gamma}+O[\gamma]^2, \\
&I_\gamma=I^{EXX}_\gamma+O[\gamma]^2,
\end{align}
and, therefore,
\begin{equation}
I_\gamma-(-\epsilon_{HOMO,\gamma})= I^{EXX}_\gamma-(-\epsilon^{EXX}_{HOMO,\gamma})+O[\gamma]^2.
\label{IE}
\end{equation}
If the LHS of Eq.~(\ref{IE}) were zero for all $\gamma$ (the IP-theorem within exact DFT), then it should hold that
\begin{equation}
I^{EXX}_\gamma- (-\epsilon^{EXX}_{HOMO,\gamma})=-\Delta \omega_{HOMO,\gamma}=O[\gamma]^2,
\label{IE1}
\end{equation}
where $\Delta \omega$ is given by Eq.~(\ref{def3}).
The fact that Eq.~(\ref{IE1}) is in contradiction to our results for Q2DEG with one miniband would be already clear unless one complication: The quantities in Eq.~(\ref{IE1}) must be calculated consistently with the orbitals at each value of $\gamma$. We have, therefore, conducted the corresponding calculations with results presented in Fig~\ref{gamma}.
The {\em linear}, rather than quadratic, scaling of  $\Delta\omega_\gamma$ with $\gamma$ at small $\gamma$ is amply evidenced by this figure. This concludes our demonstration of the violation of the IP-theorem for Q2DEG with one filled miniband not only in the EXX theory, but within the exact DFT as well.

\section{Choice of the arbitrary constant in EXX potential}
\label{AP3}

It is known that, for closed systems, $v_{xc}(\rv)$ is defined up to the addition of an arbitrary constant, while, within the open-systems formalism, this constant is fixed by the condition $v_{xc}(\infty)=0$, provided energy is measured relative to the vacuum level \cite{Levy-84}.
We note that IP of Eq. (\ref{II}) is invariant under the transformation $v_x(\rv) \to v_x(\rv)+const$, this transformation leading to 
$\epsilon_i \to \epsilon_i+const$, which is exactly compensated by the change in $\Delta \omega_i$ according to  Eq. (\ref{def3}). This is consistent with IP being a physical (observable) quantity.  
$v_x(z)$ of Eq.~(\ref{main152}) having the asymptotic form of $-\frac{1}{|z|}$ at $z\to \pm \infty$ \cite{Nazarov-16-2}, we explicitly satisfy the $v_x(\infty)=0$ condition, thus ensuring $\epsilon_{HOMO}$ to be measured relative to the vacuum level.

\end{document}